\documentclass[conference]{IEEEtran}
\usepackage{cite}
\usepackage{amsmath,amssymb,amsfonts}
\usepackage{algorithmic}
\usepackage{graphicx}
\usepackage{textcomp}
\usepackage{xcolor}
\usepackage[hyphens]{url}
\usepackage{tikz}
\usepackage{hyperref}

\usepackage{pifont}

\def\BibTeX{{\rm B\kern-.05em{\sc i\kern-.025em b}\kern-.08em
    T\kern-.1667em\lower.7ex\hbox{E}\kern-.125emX}}

 \title{ReuseSense: With Great Reuse\\  Comes   
  Greater Efficiency\\
  \Large{Effectively Employing Computation Reuse on General-Purpose CPUs
  }}
\author{\IEEEauthorblockN{Nitesh Narayana GS,
Marc Ordoñez,
Lokananda Hari,
Franyell Silfa and
Antonio González}
\IEEEauthorblockA{\textit{Department of Computer Architecture} \\
\textit{Universitat Politècnica de Catalunya,
Barcelona, Spain}\\
\{nitesh, mordonez, hari, fsilfa, antonio\}@ac.upc.edu  }
}

\begin{document}
\maketitle
\thispagestyle{plain}
\pagestyle{plain}

%%%%%% -- PAPER CONTENT STARTS-- %%%%%%%%

\begin{abstract}

Deep Neural Networks (DNNs) are the de facto algorithm for tackling
cognitive tasks in real-world applications such as speech recognition and natural language processing. DNN inference comprises numerous dot product operations between inputs and weights that require numerous multiplications and memory accesses, which hinder their performance and energy consumption when evaluated in modern CPUs.
In this work, we leverage the high degree of similarity between consecutive inputs in different DNN layers to improve the performance and energy efficiency of DNN inference on CPUs. To this end, we propose ReuseSense, a new hardware scheme that includes  ReuseSensor, an engine to efficiently
generate the compute and load instructions needed to evaluate a DNN layer accordingly when sensing similar inputs. By intelligently reusing previously computed product values, ReuseSense allows bypassing computations when encountering input values identical to previous ones. Additionally, it efficiently avoids redundant loads by skipping weight loads associated with the bypassed dot product computations.

Our experiments show that ReuseSense achieves an 8x speedup in performance and a 74\% reduction in total energy consumption across several DNNs on average over the baseline.

\end{abstract}

\section{Introduction}
\label{sec:intro}

Deep neural networks (DNNs) have rapidly gained prominence due to their ability to learn complex patterns from large datasets, leading to remarkable performance in many cognitive tasks such as image processing and computer vision (3D UNet~\cite{3dUnet}, ResNet50~\cite{resnet50}), natural language processing (BERT~\cite{devlin2018bert}), and speech recognition (DeepSpeech2~\cite{deepspeech2}). Their widespread adoption has made them ubiquitous in various platforms, ranging from high-end GPUs~\cite{a100,Yu_2023_CVPR,LightSeq2} to tiny IoT devices~\cite{xpulpnn,lai2018cmsis,tensorflowlite}. However, due to their large number of parameters and computations, DNNs have a high demand for computational power and memory requirements, which poses a significant challenge in energy consumption.
As a result, several specialized architectures (i.e., accelerators)~\cite{chen2016eyeriss,epur,brainwave2018,jouppi2023tpu,realtimednnp,nvdla} and dedicated hardware structures in CPUs~\cite{sparCE} have been proposed for improving the performance and energy efficiency of DNNs. Besides, a variety of techniques such as pruning~\cite{han2016deep,rieraprunning,zhang2016cambricon}, quantization~\cite{wu2020integer,banner2018scalable} and avoiding zero computations~\cite{ZComp,smart-DNN,SAVE} have been proposed to improve their performance and reduce their memory footprint.

CPUs are commonly employed for DNN inference due to their wide availability, cost-effectiveness, integration with existing systems, power efficiency, and scalability \cite{DNNForCPUSurvey,hazelwood2018applied}. Therefore, efficient DNN inference on CPUs has become crucial. However, 
this is challenging due to the profusion of memory access, high memory bandwidth requirements, and limited throughput of the Vector Processing Units (VPUs) in CPUs. These challenges manifest in high-energy consumption and limited performance~\cite{benchmarkingVisionKernels}, and they are even more critical on mobile and embedded CPUs with tightly constrained energy and hardware resources. Hence, in this work, we aim to improve the performance and energy efficiency of performing DNN inference on General-Purpose CPUs.

An approach to improve performance and reduce energy consumption is to reuse previously computed values to avoid some of the required memory accesses and computations. 
In this regard, some previous works~\cite{inputSimilarityMarc,UCNN,CREW,fuzzymemo} tailored to accelerators exploit the observation that for many neurons, their input values are equal for consecutive evaluations (input similarity). 
In these works, input similarity is applied to DNNs that process input sequences (e.g., audio or video); however, we have also observed a high degree of input similarity for DNNs that do not process input sequences. For instance, we have seen that for Resnet50~\cite{resnet50}, the average input similarity when processing unrelated images is 41\%. Hence, inspired by these observations, we aim to leverage input similarity to reduce the number of memory accesses and computations during DNN inference on CPUs.    

Exploiting input similarity on accelerators through hardware extensions has provided significant performance and energy efficiency improvements. For instance, the scheme proposed in \cite{inputSimilarityMarc} exploits input similarity by first caching the inputs and outputs of any given layer each time it is executed. Then, when evaluating the next set of inputs, the previous and current input elements are compared, and the new outputs are computed by adjusting the previous outputs. Regardless of the simplicity of such a scheme when implemented in a specialized accelerator, deploying such a scheme on general-purpose CPUs becomes extremely challenging. DNNs on CPUs are usually evaluated using the VPU, and thus, it is difficult to reuse single computations since VPUs process all their lanes in tandem.
Moreover, to reuse a previously computed output value, we first must compare the previous and current inputs to check if they are equal. This comparison adds overhead due to the extra branching and bookkeeping of the instructions needed to compute the similarity and skip some computations. For example, we implemented the reuse scheme proposed in ~\cite{inputSimilarityMarc} for a fully-connected layer on a state-of-the-art CPU, and when the input similarity is 45\% (typical value observed in many DNNs), the result is a slowdown of 9.7\%.

Furthermore, implementing this scheme directly on general-purpose CPUs remains inefficient due to the substantial on-chip memory requirements needed to cache previous inputs and outputs effectively. Note that this reuse scheme also faces challenges due to speculative execution in modern microprocessors. Even though the scheme employs branch instructions to skip some loads and computations, speculative execution can still execute these instructions until the branch is resolved, leading to performance and energy consumption inefficiencies. 

Since a software-based reuse scheme on CPUs is not beneficial in spite of the fact that the potential for improving performance and energy efficiency using input similarity is significant, we propose \textbf{ReuseSense}, a new hardware scheme that exploits the high degree of input similarity across different layers of a DNN. To this end, it leverages the capabilities of ReuseSensor,  a dedicated hardware unit that generates effectual load and compute instructions by sensing similar input values and directly sends the generated instructions to the backend of the CPU pipeline. Specifically, for each DNN layer, ReuseSense caches its inputs and outputs. 
Then, when computing the same layer again, the new outputs are computed by adjusting the previous outputs by the delta of dot products between corresponding inputs and weights. When an input is identical to its previous value, no operation is required, and its corresponding weights are not needed, thus saving the loading of the weights and the associated computations 
(dot products).
By efficiently skipping weight loads and bypassing computations, ReuseSense optimizes memory accesses and decreases energy consumption.

We comprehensively evaluate ReuseSense to demonstrate its effectiveness. In this regard, we assess its impact on performance and energy consumption. Our experimental results show that, on average, ReuseSense improves performance by 8x while reducing total energy consumption by 74\% with minimal hardware overhead.
In summary, the main contributions of this paper are the following:
\begin{itemize}
\item We evaluate input similarity for various modern networks and show that input similarity is also present in non-sequence based applications. 
        Previous works have demonstrated the existence of input similarity but only for sequence-based applications such as video or audio processing.
    \item We demonstrate that a software-only approach is inefficient for exploiting input similarity and reusing computations on CPUs. To address this limitation, we propose \textbf{ReuseSense}, a hardware scheme that efficiently exploits input similarity and deploys computation reuse on CPUs.

    \item  We implement our scheme on top of a state-of-the-art ARM CPU. Our experimental results show that ReuseSense improves performance by 8x on average while reducing total energy consumption by 74\%.
\end{itemize}
The rest of this paper is organized as follows: Section \ref{sec:BG} provides background on DNN inference in CPUs and the required computations and machine-level instructions. Section \ref{sec:motiv} presents the motivation and challenges of implementing the reuse scheme on CPUs. Section \ref{sec:RS}  details ReuseSense and how it leverages input similarity to efficiently exploit computation reuse on CPUs. Section  \ref{sec:exp} outlines the experimental methodology. Section \ref{sec:ev}  discusses the experimental results. Finally, Sections \ref{sec:related} and \ref{sec:conc}  
present the related work and main conclusions of this work, respectively.

\section{Background}
\label{sec:BG}

In this section, we present an overview of the computations involved in Deep Neural Networks (DNNs) and introduce the concept of exploiting input similarity to achieve computation reuse. Also, we delve into the functioning of the principal assembly instructions employed to implement the CPU kernels for DNN evaluation.

\subsection{DNN Computations }

DNNs are the core algorithm for machine learning applications. They consist of several layers stacked on top of each other. The two most commonly employed layers in DNNs are Fully-Connected (dense) and Convolutional Layers. Although these two layers are conceptually different, their calculations mainly involve matrix multiplications. Fully-connected layers are typically computed using General Matrix Multiplications (GEMMs) for batch sizes greater than one, and Vector-Matrix Multiplications for batch sizes equal to one. In contrast, Convolutional layers are evaluated as either a single large GEMM or a series of smaller GEMMs, depending on the specific implementation \cite{anatomyCNN}. Notably, convolutions can extend to multiple channels, each representing a GEMM. Furthermore, GEMMs are eventually translated into a series of Vector-Matrix Multiplications. As a result, Vector-Matrix multiplications become the predominant operation in DNNs.  
The following equation mathematically represents the dot product operation used for Vector-Matrix multiplication. 

\begin{equation}
    O = \sum_{i=1}^{n} w_i \cdot x_i \label{eqn:vec-mat}
\end{equation}
where $w$ is a vector containing the synaptic weights, $x$ is the input vector, and $O$ is an element of the output features.

CPUs exploit the inherent parallelism in Vector-Matrix multiplication through their built-in vector units. In this regard, various software stacks~\cite{intelmlk,amdoc} have been introduced to ensure that DNNs deployed on VPUs exploit this parallelism efficiently. Particularly, ARM platforms employ ARMNN~\cite{arm_nn} as a software stack that optimizes DNNs based on their requirements and the underlying hardware. In this work, we use an ARM CPU as the baseline and employ the ARMNN framework to implement the DNN models used in our evaluations. Nonetheless, DNN kernels implemented for other ISAs commonly employ similar vector instructions, and thus, our analysis and conclusions can be applied to other ISAs. Also, DNN models are usually quantized 
to increase performance, and in this work, we employ quantized DNNs (i.e., 8-bits) for evaluations. 
% to increase performance, and hence, we employ DNNs quantized in 8-bits for evaluations. 

\subsection{Similarity and Reuse}

\begin{figure}

    \centering
	\begin{align}
	 \Delta &= I_c - I_{p}
	\label{e:delta}
	\end{align}
\begin{align}
	   O_c &= O_p + (I_{p}- I_c)*w
	\label{e:current_output_red}
	\end{align}
 \begin{align}
	  O_c = O_p + \Delta \cdot w
	\label{e:current_output}
	\end{align} 
	\caption{Dot product computation for an output feature ($O_c$) based on its previous output ($O_p$) and the difference between the current and previous input vectors, $I_c$ and $I_p$, respectively.}
	\label{f:reuse_equations}
\end{figure}
Previous work \cite{inputSimilarityMarc} observed that the inputs to any given layer of a DNN exhibit a significant degree of similarity when the DNN is used for sequence-processing applications such as video and language processing. In this regard, \textit{similarity} is defined as the percentage of identical values between two consecutive inputs for a given layer. To exploit this observation, they expressed the dot product computation in Equation~\ref{eqn:vec-mat} as a function of a previous computation and a previous input as shown in Equations~\ref{e:delta}-\ref{e:current_output}. For these equations, $w$ is a vector of weights, whereas the vector $\Delta$ represents the difference between two input vectors: the current input vector $I_c$ and the previous input vector $I_{p}$. Similarly, $O_p$ represents a previous dot product computation, and $O_c$ corresponds to the computed dot product. Note that in Equation~\ref{e:current_output}, if any element of the vector $\Delta$ is 0, indicating no difference between two elements of the input vectors, the corresponding multiplication and addition can be skipped.

\subsection{Dot Product Instructions for Vector-Matrix Multiplication}

For evaluating DNNs on ARM platforms, the ARMNN framework is normally used. This highly optimized framework employs customized kernels from the ARM Compute Library (ARMCL)~\cite{arm_CL}. ARMCL aims to maximize the overall utilization of the vector unit for GEMM computation and neural network processing. To this end, the kernels provided by ARMCL are implemented using the \texttt{sdot} and \texttt{mla} instructions from the ARM Scalable Vector Extension (SVE)~\cite{ARM-SVE}. SVE instructions provide support for byte, half-word, word, and double-word encodings, which allows accessing vector registers at different levels of granularity and supporting DNN models quantized for different precisions.

\begin{figure}[h]
  \centering
  \includegraphics[width=1\linewidth]{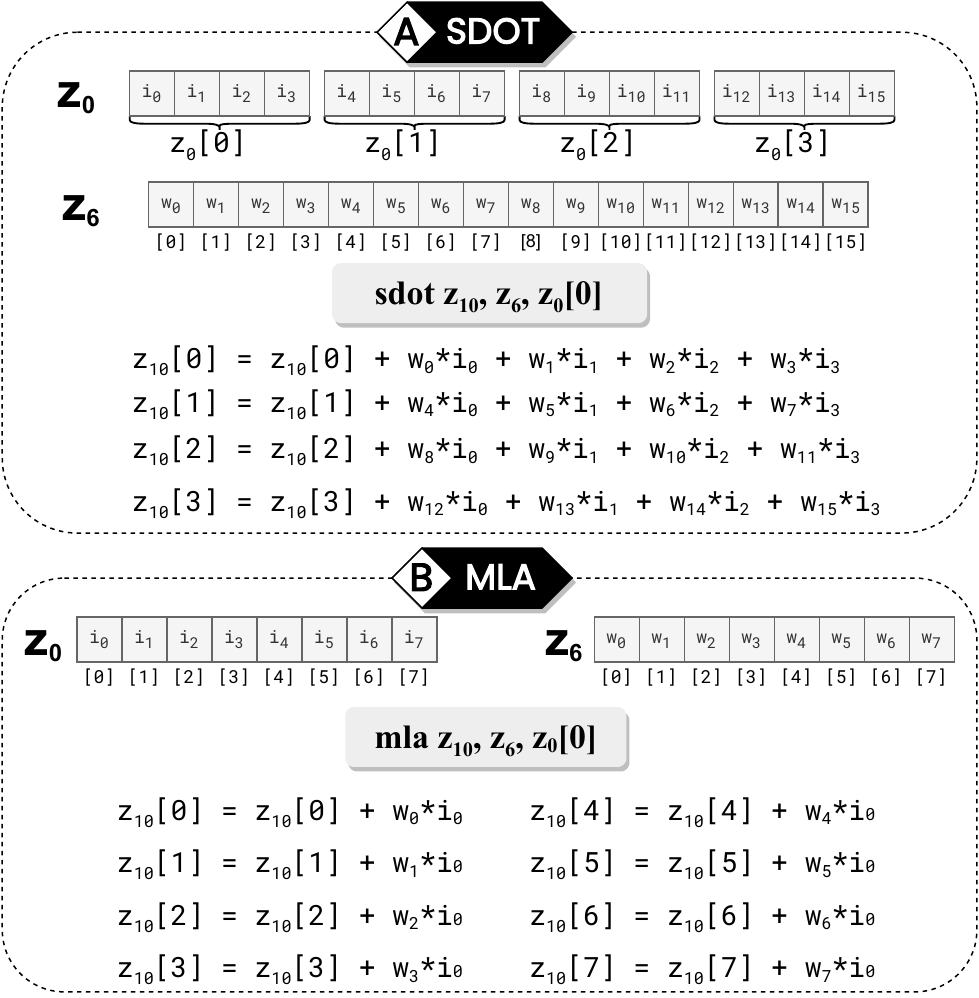}
  \caption{ \texttt{sdot} and \texttt{mla} instructions in ARM SVE. This configuration is for a 128-bit vector length. \texttt{z10} is the destination register and \texttt{z6}, \texttt{z0} are the source vector registers, respectively.}
  \label{fig:sdot-mla}
\end{figure}

 Consider the \texttt{sdot} instruction: \textit{sdot dst,src2,src1[k]}. It takes two source vector registers as input (i.e., \texttt{src1} and \texttt{src2})  and a vector destination register (i.e., \texttt{dst}). For this instruction, each source vector register consists of $N$ sub-vectors containing $M$ elements. Then, when executed, the instruction performs a dot product operation between each of the sub-vectors of the second source register (\texttt{src2}) and one of the sub-vectors of the first source register (\texttt{src1[k]}, specifies sub-vector $k$ from \texttt{src1}). Moreover, each dot product result is accumulated with the destination register (\texttt{dst}). Note that the destination register is divided into $N$ accumulators. For example, conceptually, the \texttt{sdot} instruction shown in Fig. \ref{fig:sdot-mla}-A works as follows for vector registers of 128-bits. First, each source vector register is divided into four sub-vectors containing four signed 8-bit integer values. In this example, sub-vector zero from the register \texttt{z0} contains the values $i_0$ to $i_3$, whereas the register \texttt{z6} is divided into four sub-vector containing the elements $w_0$ to $w_3$, $w_4$ to $w_7$, and so on. Then, a dot-product between sub-vector zero from register \texttt{z0} and each sub-vector of \texttt{z6} is performed. Finally, 
 the intermediate results of these dot products are accumulated in the register \texttt{z10}, which in this example is divided into four accumulators.

Consider the \texttt{mla} instruction: \textit{mla dst,src2,src1[k]}, shown in Fig.~\ref{fig:sdot-mla}-B. This instruction operates on two source vector registers as input and a vector destination register. Each source vector register consists of n-bit integers. The instruction performs an element-wise multiplication between all the elements of \texttt{src2} and an element of \texttt{src1} specified by the index $k$, then it adds the results to the corresponding element in the destination register. For example, in Fig.~\ref{fig:sdot-mla}-B for 128-bit vector registers and 16-bit elements, the instruction performs the multiplication between the eight elements of \texttt{z6} and element 0 of \texttt{z0}. Then, the multiplication results are accumulated with the eight elements of \texttt{z10}.  

Note that there are other variants of these instructions, and for details on those, we refer the reader to the Arm A-profile A64 ISA Documentation \cite{ARMv9}.

\section{Motivation}
\label{sec:motiv}

\subsection{Input Similarity}

In prior works~\cite{inputSimilarityMarc,deepReuse,fuzzymemo}, input similarity has been primarily studied in DNNs for sequence-processing applications, where inputs correspond to consecutive elements such as frames of videos or sound. However, in this work, we also aim to exploit input similarity in networks where inputs are unrelated (i.e., networks where the current input is not correlated with the previous input). We show the existence of input similarity despite the absence of temporal dependencies. 

\begin{figure}
  \centering
  \includegraphics[width=1\linewidth]{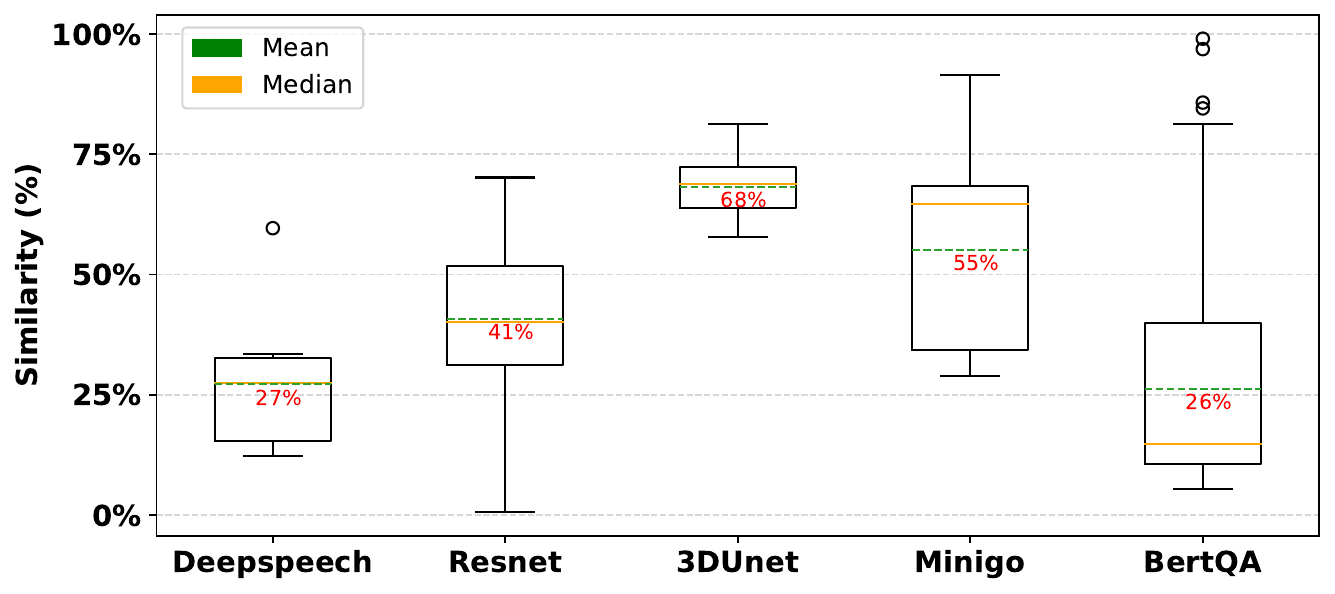}
  \caption{Average input similarity across layers for different DNNs.}
  \label{fig:similarity}
\end{figure}

We conducted the following experiment to measure the input similarity for several DNNs (described in Section~\ref{sec:exp}). First, for DNNs without correlated inputs, we select random inputs from their respective datasets. Then,  we compute the input similarity for each layer by comparing its current input with the one in the previous evaluation. For sequence-processing DNNs, we measure the similarity by comparing consecutive timesteps (i.e., consecutive audio frames) in the current input.

Fig.~\ref{fig:similarity} shows a distribution of the average input similarity for the layers on different DNNs. For sequence-processing DNNs (Deepspeech, 3DUnet, Minigo), the average input similarity ranges from 27\% to 68\%. This result is consistent with previous works that pointed out this phenomenon.
On the other hand, the input similarity for networks with no correlated inputs (Resnet, BertQA) is also surprisingly high, ranging from 25\% to 41\%. Furthermore, for some layers, the input similarity is greater than 80\%, as evidenced by the outliers depicted in  Fig.~\ref{fig:similarity}.

\begin{figure}[]
  \centering
  \includegraphics[width=1\linewidth]{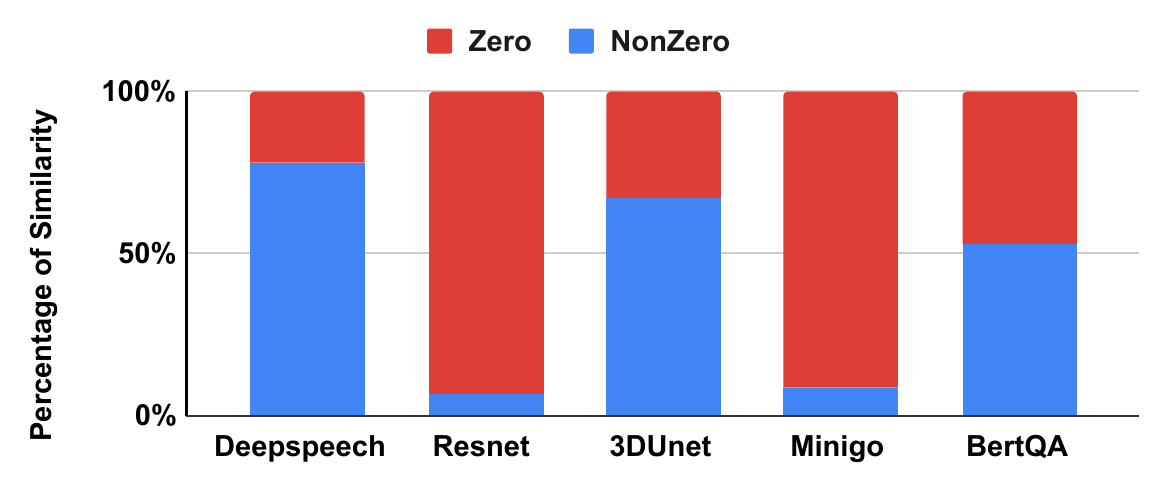}
  \caption{Breakdown of input similarity from consecutive input values that are zeros or nonzeros but identical. }
  \label{fig:simDist}
\end{figure}

The sources for the input similarity come from two scenarios: when two consecutive inputs contain values that are identical and nonzero or when they are zero. Fig.~\ref{fig:simDist} illustrates the distribution of input similarity based on these criteria. For some networks, the distribution of nonzero similar values spans from 50\% to 75\% of the overall similarity. In contrast, for other networks, over 90\% of the overall input similarity arises from zero values. This phenomenon can be attributed to the combined influence of using low precision (i.e., 8-bit quantization) and the occurrence of activation functions, such as ReLU.

\subsection{Challenges for Reusing Computations in CPUs}

\begin{figure*}
  \centering
  \includegraphics[width=1\linewidth]{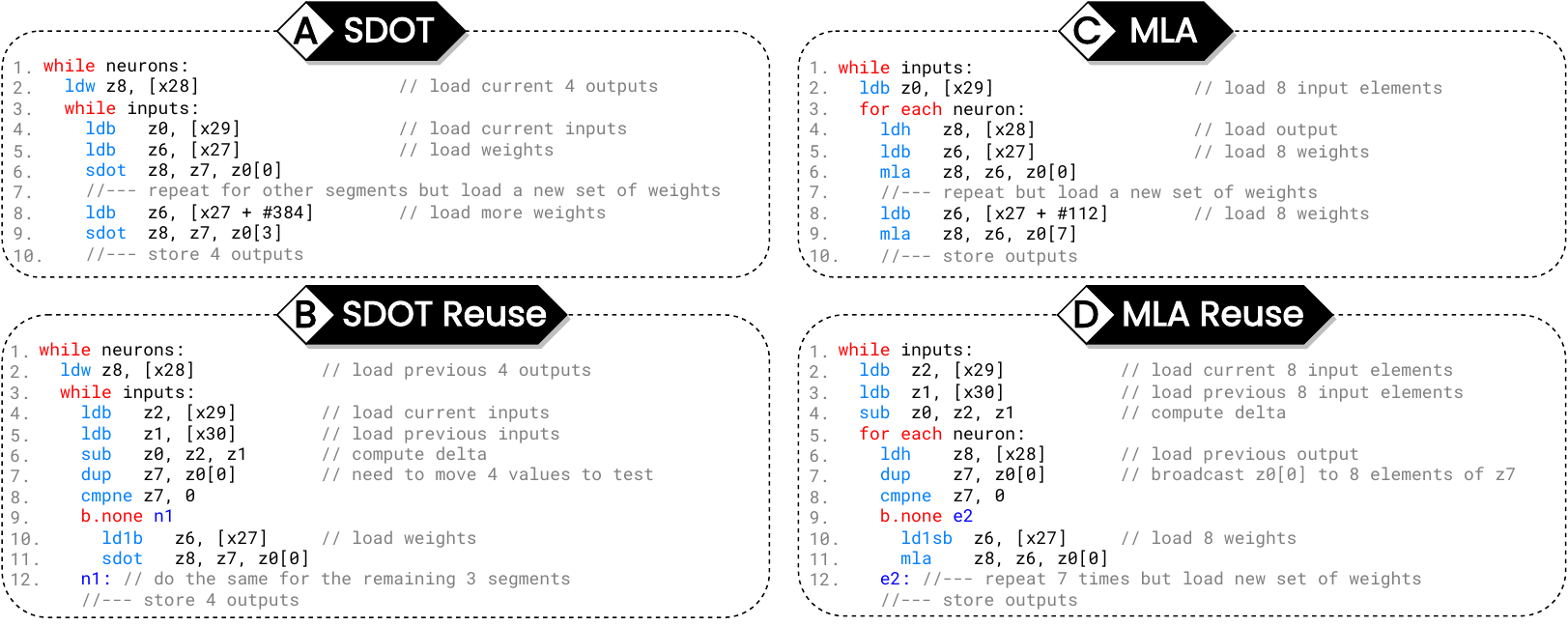}

  \caption{Kernel Pseudo-code for performing vector-matrix multiplications on CPUs based on the \texttt{sdot} and \texttt{mla} instructions. A and C do not employ computation reuse, whereas B and D do.}
  \label{fig:softReuse}
\end{figure*}

Aiming to exploit input similarity and employ computation reuse in CPUs, we modified an ARMNN kernel for evaluating DNN layers. In the rest of this section, we first explain a typical DNN kernel.
Then, we detail the modifications done to exploit input similarity in software and explain the challenges and inefficiencies of this software-based scheme for computation reuse.

Figs.~\ref{fig:softReuse} A and C display a pseudo-code of a typical vector-matrix multiplication ARMNN kernel used to compute DNN layers based on the \texttt{sdot} and \texttt{mla} instructions, respectively. In essence, the \texttt{sdot} and \texttt{mla} kernels compute and accumulate the partial dot products for a given input vector and weight matrix. Note that in these pseudo-codes, we refer to the computation of a dot product as computing the output of a neuron. Also, to evaluate a neuron the input vector is divided into sub-vectors to fit in the vector registers of the VPU.

\begin{figure}
  \centering
  \includegraphics[width=1\linewidth]{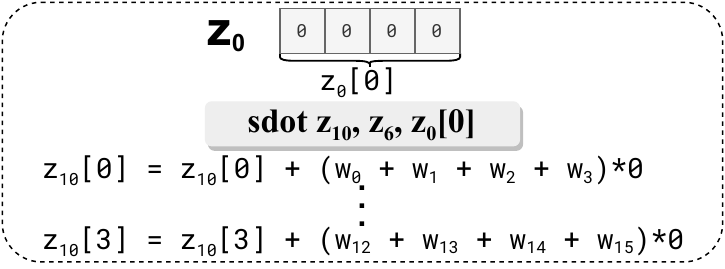}
  \caption{All the elements in the sub-vector 0 (\texttt{z0[0]}) must be zero to make this \texttt{sdot} instruction ineffectual.}
  \label{fig:sdot-delta}
\end{figure}

To employ computation reuse in the \texttt{sdot} based kernel, shown in Fig.~\ref{fig:softReuse}-A,
we re-implemented it according to Eqns~\ref{e:delta}-\ref{e:current_output}. The modified version is shown in Fig.~\ref{fig:softReuse}-B, and it aims to skip the compute instruction when the current input values are identical to previous input values by using branch instructions. Also, it avoids loading the weights associated with the compute instruction. To this end, first, it loads the previous inputs and output for each neuron. Then, it computes the delta between the current and previous input vectors, following the equations defined in Eqns~\ref{e:delta}-\ref{e:current_output}.
Since the \texttt{sdot} instruction computes dot products between sub-vectors (Fig.~\ref{fig:sdot-mla}), the kernel first checks if all the deltas in one of the sub-vectors are equal to zero. Note that computations and loading the weights can only be skipped when all the deltas in a given sub-vector are zero, as illustrated in Fig. \ref{fig:sdot-delta}. Subsequently, if all the deltas in the sub-vector are zero, the kernel skips the process of loading weights and computing those sub-vectors. However, if any delta is nonzero, the kernel has to proceed with those evaluations. This process is repeated for other sub-vectors until all the inputs are processed for all the neurons.

The modified \texttt{sdot} kernel encounters challenges in efficiently employing computation reuse for two primary reasons. Firstly, all the deltas in a given sub-vector must be zero to avoid loading weights and skipping computations. Experimentally, we observed that this rarely occurs. For instance, in Resnet, such cases account for only 13.9\% of the overall network similarity.
Secondly, modern processors utilize speculative execution, which means that even if the branch outcome is to skip loading and computation, the processor may speculatively execute the loads and computation instructions until the branch is resolved. This speculative execution prevents the goal of avoiding unnecessary work.

In a similar vein, the adapted \texttt{mla} kernel, as illustrated in Fig.~\ref{fig:softReuse}-D, has been tailored to support computation reuse and avoid ineffectual instructions, akin to the \texttt{sdot} reuse kernel. In this regard, the 
delta values are computed similarly; however, 
after computing the delta values for current and previous input vectors, an indexed scalar value ($k$) from it (i.e., \texttt{z0[k]}) is copied into a separate register.
Then, this scalar value is multiplied by a set of weights using a \texttt{mla} instruction. Note that if the indexed delta value is zero, these computations can be avoided, similar to the \texttt{sdot} reuse kernel, thus eliminating the need to load corresponding weights.  
However, it should be noted that, like the \texttt{sdot} kernel, this modified kernel also faces challenges due to the speculative execution nature of modern processors, which can impact the efficacy of computation reuse.

Experiments reveal that compared to the baseline kernel in Fig.~\ref{fig:softReuse}-A, the kernels in B, C, and D exhibit slowdowns of 10\%, 34\%, and 31\% in execution time, respectively.
In other words, the \texttt{sdot} based kernel outperforms the mla-based kernel, but more importantly, none of the two versions can leverage the potential benefits of computation reuse as a software-only approach. Hence, motivated by the large percentage of input similarity found in DNNs and the challenges to leveraging this in a software-based reuse approach, we propose ReuseSense, a hardware scheme that aims to mitigate these challenges by generating and feeding the kernel instructions directly to the backend of the CPU's pipeline while skipping the ineffectual instructions due to computation reuse and input similarity.

\section{ReuseSense}

\begin{figure}[]
  \centering
  \includegraphics[width=1\linewidth]{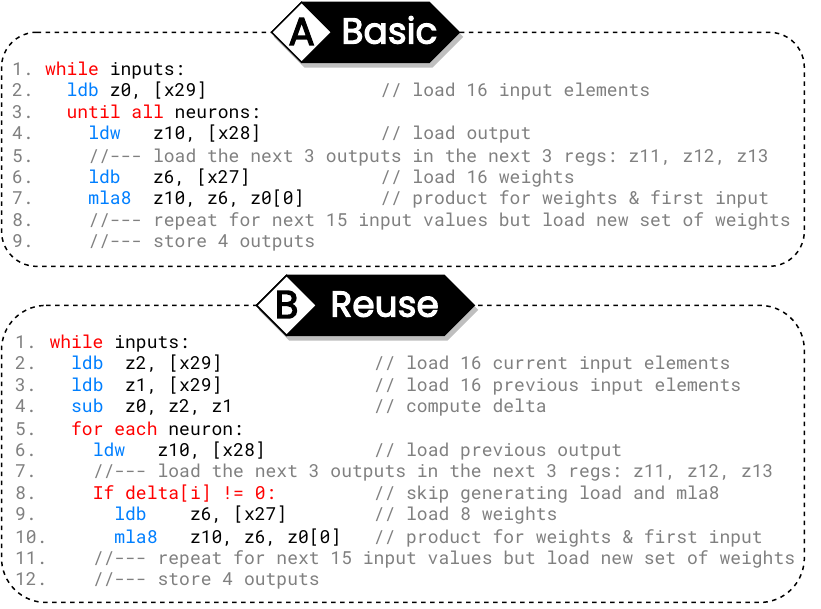}
  \caption{Kernel Pseudo-code used by ReuseSensor to generate instructions. }
  \label{fig:CIKernels}
\end{figure}

The primary objective of ReuseSense is to obviate the execution of ineffectual instructions that result from leveraging input similarity for computation reuse. To achieve this, ReuseSense employs ReuseSensor, a hardware component responsible for generating the instructions needed to evaluate a DNN layer. To this end, the ReuseSensor generates instructions based on the kernels depicted in Fig.~\ref{fig:CIKernels}, and it also transmits the generated instructions directly to the backend of the CPU's pipeline for further processing.
Note that in Fig.~\ref{fig:CIKernels}, the basic kernel (Fig.~\ref{fig:CIKernels}-A), is employed to evaluate 
DNN layers without reusing previous computations, whereas reuse kernel (Fig.~\ref{fig:CIKernels}-B) employs reuse. Also, both kernels employ the \texttt{mla8} instruction to perform the multiplication and accumulation of weights by inputs when evaluating a DNN layer. In the rest
of this section, we describe how to use ReuseSense from a programmer's perspective and present the inner workings of ReuseSensor and \texttt{mla8}.

In order to employ ReuseSense to evaluate a given DNN layer, the programmer must utilize a new instruction that we call CallReuseSensor (CRS). The CRS instruction is of the form \texttt{crs src}, where it takes a scalar source register (\texttt{src}) that contains the base address of a structure containing the kernel parameters needed to compute one of the kernels shown in Fig.~\ref{fig:CIKernels}. This structure incorporates the input length, output length, input address, weight address, output address, and the address of previous inputs. Also, it includes a flag (kernelMode) to indicate if the unit is reusing computations. Furthermore, this structure contains a parameter to indicate if the instructions are generated following an Input Stationary or Output Stationary dataflow.   

When leveraging ReuseSense to evaluate a DNN model within a given ML framework (i.e., ARMNN), the framework must evaluate each layer in sequence. Moreover, we assume that the ML framework performs any rearrangements of the weights and tiling when required. Then, to evaluate a given layer or tile, the framework must invoke the \texttt{crs} instruction. Note that before calling this instruction, the underlying framework must set up the structure containing the parameters that will be passed to it. Moreover, we leverage the underlying framework to update the data for previous inputs and outputs upon completing the execution of a \texttt{crs} instruction. After this, the process can be iteratively applied to compute all the layers in a given DNN model as required.

\label{sec:RS}

\subsection{MLA8 Extension}

In the preceding section, we highlighted the challenges of exploiting input similarity in software. Notably, the \texttt{sdot} instructions require all delta values in a sub-vector to be zero simultaneously to skip a computation. Furthermore, the \texttt{mla} instructions typically only handle 16-bit elements or bigger. This presents an issue when evaluating quantized DNNs since lower precisions (i.e., 8 bits) are commonly employed, and as a result, the VPU lanes are not fully utilized. Also, accumulating multiplications of 16-bit elements on a 16-bit register can result in an overflow.

Note that when employed in a reuse-based kernel (Fig.~\ref{fig:CIKernels}-B), the \texttt{mla} instruction offers the capability to be skipped, even if just one delta value equals zero. This relaxation of constraints distinguishes \texttt{mla} from the \texttt{sdot} instruction, where all deltas in a sub-vector are required to be zero in order for the computations and loads to be skipped (Fig \ref{fig:sdot-delta}). Hence, since \texttt{mla} is
amenable for exploiting input similarity, we used it to implement our reuse scheme in hardware. However, we first extended \texttt{mla} to mitigate its drawbacks. We call this extension \texttt{mla8}.

\begin{figure}
  \centering
  \includegraphics[width=1\linewidth]{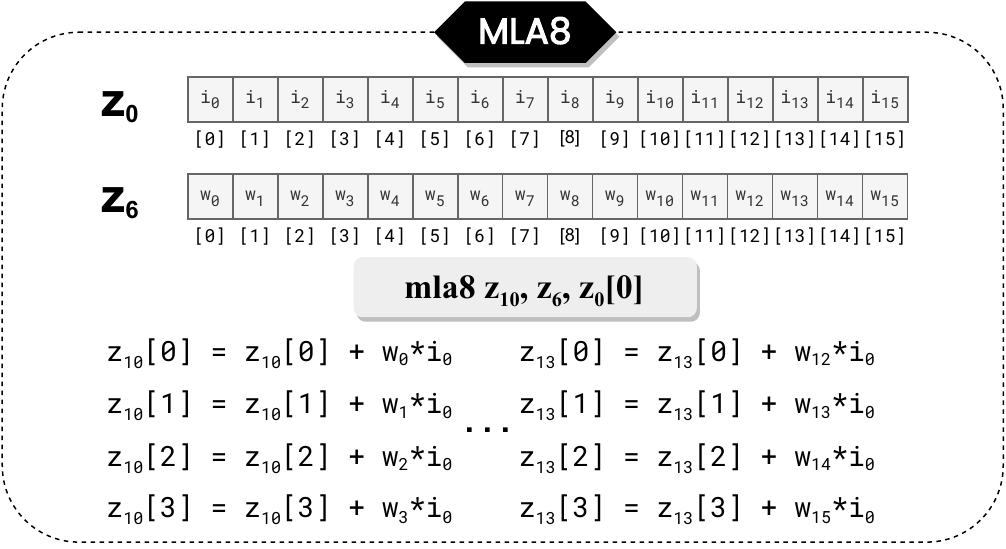}
  \caption{Structure of \texttt{mla8}. The instruction uses destination registers \texttt{z10} through \texttt{z13}. The representation is for a 128-bit vector length configuration.} 
  \label{fig:mla8}
\end{figure}

As depicted in Fig. \ref{fig:mla8}, the \texttt{mla8} instruction takes two input source vector registers, where each register consists of a series of signed 8-bit integers. During computation, each 8-bit integer in the first source register (e.g., \texttt{z6[0]} through \texttt{z6[15]}) is multiplied by the specified element in the second source register (e.g., \texttt{z0[0]}). The resulting products are then accumulated in the corresponding destination register, as shown in Fig.~\ref{fig:mla8}. 
Unlike the previous \texttt{mla} version, \texttt{mla8} requires more destination registers (i.e.,  four 128-bit vector registers for destination registers). The main reason is that the result of each multiplication is accumulated using 32-bits. Hence, using four destination registers enables the \texttt{mla8} instruction to store all products from the two input source registers, providing a more efficient and flexible solution than the base \texttt{mla} instruction. 

\subsection{ ReuseSensor}

\begin{figure}[]
  \centering
  \includegraphics[width=1\linewidth]{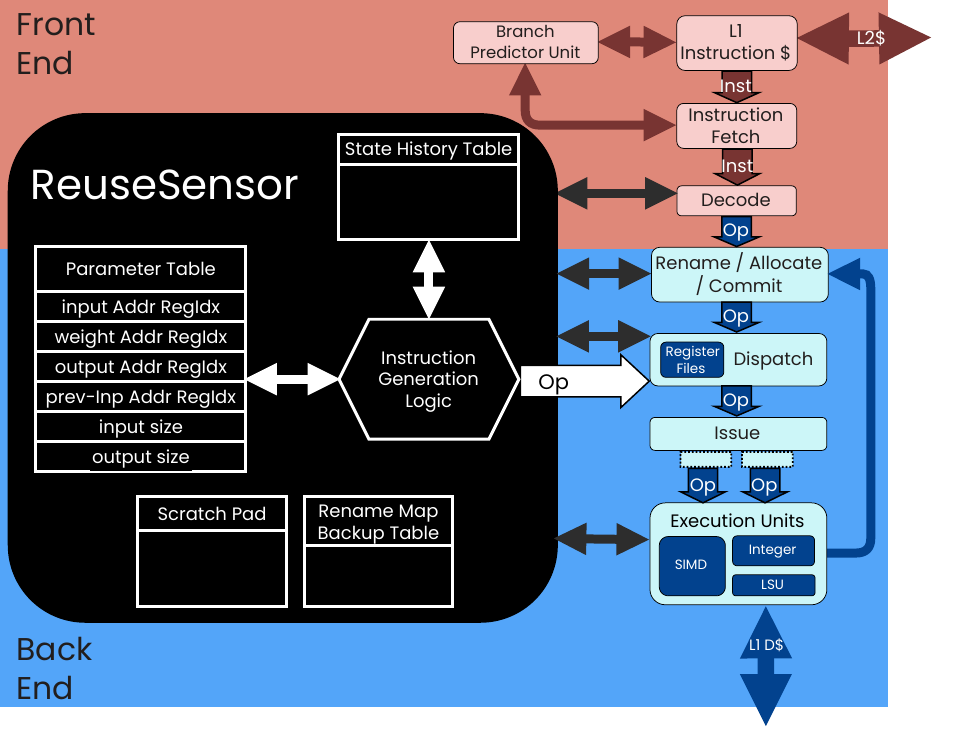}
  \caption{CPU Microarchitecture diagram with ReuseSensor. }
  \label{fig:pipeCI}
\end{figure}

Based on the kernels in Fig.~\ref{fig:CIKernels}, ReuseSense controls and generates which instructions are fed to the backend of the pipeline, notably at the dispatch stage. By strategically positioning the unit between the front-end and back-end of the pipeline (as depicted in Fig. \ref{fig:pipeCI}), only the most relevant and necessary instructions are created, while ineffectual computations and the weight loads associated to those compuations are skipped. Also, it eliminates the challenges faced due to the speculative execution nature of the CPU (Section \ref{sec:motiv}).

Illustrated in Fig. \ref{fig:pipeCI}, ReuseSensor comprises several integral components. First, it employs an on-chip scratchpad memory to store register values from the vector register file. Additionally, a Rename Map Backup table is utilized for managing rename mappings of registers used by the unit. The parameter table is responsible for storing the kernel parameters described earlier. It also features an instruction generation logic responsible for generating and skipping instructions, aligning with the kernels depicted in Fig. \ref{fig:CIKernels}. Lastly, a state history table retains the state of the instruction generation logic and parameter table. This table is utilized to recover from faults resulting from the instructions generated by the unit.

\subsubsection{Overview of ReuseSensor Execution} The ReuseSensor performs the following steps when evaluating a DNN layer:

\textbf{\ding{202} Preparing State}
When a CRS instruction is encountered during the decode stage, a signal is sent to activate ReuseSensor, and the decode stage stops accepting instructions from the fetch stage. ReuseSensor enters the preparing state, waiting for the pipeline to drain any remaining instructions older than CRS. During this phase, the unit starts backing up the vector physical registers into an on-chip scratchpad memory and takes a backup of the rename map table for the vector register file into the rename map backup table.

\textbf{\ding{203} Generate State}
Once all the instructions before CRS have been committed and the pipeline is drained, ReuseSensor moves to the Generate State. The instruction generation logic uses the address specified in the source register of CRS to load the required kernel parameters.  After loading them, it populates their respective metadata into the parameter table.

\textbf{\ding{204} Kernel Instruction Generation} in this state, ReuseSensor starts to generate instructions according to the kernel structure shown in Fig. \ref{fig:CIKernels}. Specifically, it generates instructions to load the current and previous inputs and the previous output. Moreover, it creates the instructions for computing the delta values. Finally, in this state, ReuseSensor decides when to skip generating weight loads and computation based on delta values
and generates them otherwise.

\textbf{\ding{205} Finishing State}
After all the instructions are generated, ReuseSensor moves to the finishing state, waiting for them to be committed.

\textbf{\ding{206} Restore State}
Finally, when all the instructions generated by ReuseSensor are committed, the unit starts restoring the backed-up vector physical registers and the vector rename map table. Subsequently, ReuseSensor unblocks the decode stage, allowing normal program execution to resume.

\subsubsection{Workflow of ReuseSensor}

To elucidate the working of ReuseSense, consider the reuse kernel in Fig.~\ref{fig:CIKernels}-B as a reference. Note that the following process also applies to the basic kernel (Fig.~\ref{fig:CIKernels}-A), but we focus on the reuse-based kernel for brevity. First, to activate ReuseSensor, the \texttt{crs} instruction is invoked. Then, in \textbf{\ding{202}}, when the ReuseSensor moves to the preparing state, it waits for the pipeline to drain any remaining instructions preceding \texttt{crs} in the program flow. By doing this, we ensure that the instructions generated by ReuseSensor are not subject to interference from any ongoing computations. Moreover, the Decode stage is temporarily blocked during this phase to avoid potential conflicts and anomalous behaviour, preventing newer instructions from progressing further into the pipeline. For instance, in scenarios where store instructions modify the kernel parameters or any data for the DNN model (i.e., weights), such interference could lead to unnecessary faults and squashes in the pipeline. Additionally, newer instructions may compete for resources like physical registers and queue entries with the instructions generated by the ReuseSensor, leading to performance degradation. By halting the Decode stage and allowing the pipeline to drain, we effectively eliminate any interference between regular program instructions and those generated by the ReuseSensor.

Our analysis demonstrates that the number of cycles required for draining the pipeline is minor compared to the total cycles during which ReuseSensor operates, even across various layers of the networks. Thus, the waiting time is minimal, making it a negligible overhead. Note that the ReuseSensor operation time spans from its activation to the commit of the last instruction it generates, and the overhead of draining the pipeline is less than 0.1\% of the ReuseSensor's operating cycles. Additionally, the current implementation of ReuseSense prevents any context switching on the CPU core where a \texttt{crs} instruction is being executed. 

Finally, during the preparing state, the ReuseSensor starts backing up the vector physical registers marked as ready in the vector register file. Then, it returns them to the free list. Also, if any vector physical register is being written, it will be backed up once it becomes ready. Hence, we ensure that ReuseSensor can access all available vector physical registers. Additionally, ReuseSensor takes a backup of the rename map table for the vector register file, which is done to restore the previous mapping of the vector rename table once the \texttt{crs} instruction is complete.

In \textbf{\ding{203}}, ReuseSensor proceeds to the instruction generation stage after the pipeline is drained. Notably, The instruction generation logic uses predefined micro operation and register encoding to generate the required instructions. First, during this stage, it 
generates instructions to load the required configuration parameters for the kernel evaluation using the source register of the \texttt{crs} instruction as the base address. To this end, ReuseSensor acquires physical registers from the free-list and uses them to generate the load instructions to fetch the configuration parameters. Then, a mapping between physical registers and configuration parameters is done and stored in the parameter table. This simple mapping enables the unit to keep track of the parameters and utilize them later in generating kernel instructions. Then, the load instructions are generated and sent to the dispatch stage, where they follow the typical pipeline flow. As the load instructions commit, the configuration parameters become known. At this point, their entries in the parameter table are updated, providing the necessary information for subsequent computations.  

In \textbf{\ding{204}}, after loading the configuration parameters, the ReuseSensor proceeds to generate instructions aligned with the kernels depicted in Fig.~\ref{fig:CIKernels}. The instruction generation logic in the unit produces enough instructions per cycle (i.e., four instructions) to keep the CPU and VPU fully utilized. Note that when generating instructions, a specific architectural register is assigned for each type of instruction. For example, the load instruction to load the input values is assigned to \texttt{z1}. ReuseSensor then chooses a physical register from the free-list and maps this fixed architectural register to the chosen physical register in the rename table. This straightforward mapping allows the instruction to pass through the backend of the pipeline just like any other decoded and renamed instruction. Note that the instruction generation logic assigns a sequence number to each generated instruction, which helps track the state of the instructions until they are committed.   

Following the pseudo-code in Fig.~\ref{fig:CIKernels}-B, when evaluating it, ReuseSense generates the instructions required for its evaluation in the following manner: First, for the instruction to load input (line \texttt{2} in Fig. \ref{fig:CIKernels}-B), the instruction generation logic procures a vector physical register from the free-list and maps it to a fixed architectural register in the vector rename map table. Simultaneously, the pipeline scoreboard is updated accordingly, and the register that holds the input address is obtained from the parameter table. Likewise, the ReuseSensor generates instructions for loading the previous input and the subtraction instruction for computing the delta (lines \texttt{3}, \texttt{4} in Fig.~\ref{fig:CIKernels}-B respectively). Next, the unit similarly generates instructions to load the previous output  (line \texttt{5} in Fig.~\ref{fig:CIKernels}-B).

Then, before loading the weights, ReuseSensor checks whether the result of the subtraction instruction (deltas) is ready. If it is not ready, the unit waits for the result. Once the subtraction instruction becomes ready to be committed, the delta values are copied into the in-unit delta value register. This register is then accessed to check which deltas are equal to zero (line \texttt{8} in Fig.~\ref{fig:CIKernels}). In case a delta value is non-zero, the instruction to load weights and the \texttt{mla8} instruction will be generated and executed; otherwise, they are skipped.
Additionally, if an overflow is detected during the delta computation, ReuseSensor addresses this by generating two separate computation instructions for the same set of weights. This process involves splitting the overflown delta into two components, ensuring it remains within the permissible range. Notably, our experiments indicate that such occurrences account for less than 0.01\% across all the networks. As a result, this method does not introduce any significant additional overhead.
The above process is repeated until all the inputs specified in the configuration parameters are evaluated.   

Each instruction generated by the ReuseSensor unit adds an entry to the state history table. This table records the current state of the instruction generation logic and parameter table at a particular moment. Also, this table is indexed using the sequence number of the instruction being generated. The purpose of maintaining this history is to serve as a mechanism for fault recovery. If a load-store reordering fault is detected due to an instruction generated by the ReuseSensor, it can refer to this table and revert to the saved state. As a part of the regular pipeline flow, the entries in the state history table are evicted each time the corresponding instruction with the same sequence number is successfully committed. Additionally, during squashes due to load-store reordering faults, the table is appropriately managed to maintain its accuracy and relevance.

In \textbf{\ding{205}}, once all the instructions for the kernel evaluation have been generated and sent to the pipeline, the ReuseSensor transitions to the finishing state, awaiting any pending instructions to be committed. Then, in \textbf{\ding{206}}, when all the pending instructions are committed, it starts restoring the vector physical registers from the scratchpad to the vector physical register file and the vector rename table from the rename map backup table. Furthermore, the ReuseSensor frees all the integer registers that it was using. Finally, after completing the restoration, the ReuseSensor unblocks the decode stage, enabling the CPU to proceed with its normal execution flow.

\section{Experimental Methodology}
\label{sec:exp}

\textbf{Workloads:} Our experiments are conducted using various DNNs which are quantized in 8-bit, as summarized in Table \ref{tab:dnnSummary}. Each network takes different types of inputs. In this regard, Resnet takes images of various categories as input and predicts their classes. 3DUnet  takes annotated volumetric medical images to provide dense 3D Tumour segmentation \cite{3dUnet}. On the other hand, BertQA takes contexts (paragraphs) and questions as input and gives the start and end index of the answer from the paragraph, which can then be converted to the actual answer. Minigo processes images that represent the positions of the stones for each color to give the next move. Finally, Deepspeech takes an audio file as input and reports a transcription. For the evaluation of the input similarity and functional analysis of the networks, we use Pytorch \cite{pytorch} and TensorFlow \cite{tensorflowlite}.
\begin{table}
\caption{DNN models used in our experiments. Similarity refers to the average percentage of input similarity.}
\centering
\begin{tabular}{|l|l|l|l|}
\hline
\textbf{Network} & \textbf{Application Domain} & \textbf{Dataset} & \textbf{Similarity} \\ \hline
BERT-QA          & Question Answering          & SQUAD            & 26\%     \\ \hline
3DUnet           & Image Segmentation          & TCGA-LGG         & 68\%   \\ \hline
ResNet50         & Image Classification        & ImageNet         & 41\%     \\ \hline
DeepSpeech2      & Speech Recognition          & LibriSpeech      & 27\%   \\ \hline
Minigo           & Reinforcement Learning      & -                & 55\%   \\ \hline
\end{tabular}
\label{tab:dnnSummary}
\end{table}

For our experiments, we employ a modified version of the Gem5 simulator \cite{gem5} with a customized configuration, as detailed in Table \ref{tab:cpuConfig}. Our simulation environment is based on the ARM Cortex-A76-like configuration \cite{ARM-SVE}, utilizing a 128-bit vector length. It is crucial to emphasize that while the current implementation utilizes ARM-ISA, ReuseSense is designed to be ISA-independent. Hence, they can be deployed on any ISA by extending respective ISA Vector extensions to support ReuseSense.

To evaluate energy consumption and hardware overhead, we employed McPat \cite{mcpat} with a 32 nm technology node. This allowed us to extract energy results by passing Gem5 statistics to McPat. For estimating the energy and area of ReuseSensor, we utilized CACTI \cite{cacti} to model the scratchpad memory. At the same time, for the remaining logic and structures, we implemented them in Verilog and obtained the relevant metrics using Synopsys Design Compiler \cite{Synopsys}.

\begin{table}[h]
\caption{Baseline CPU Configuration.}
\label{tab:cpuConfig}
\centering
\begin{tabular}{|c|c|}
\hline
\textbf{Component} & \textbf{Specification} \\ \hline
CPU (@1.5GHz) & 128 Int RF, 192 FP RF, 48x128-bit Vector RF \\
 % & Merged Register-File is assumed.\\
 & 80 IQ, 32 LQ, 48 SQ, 128 ROB Entries \\
 & 4-wide fetch/decode/rename/commit \\
 & 8-wide issue/dispatch/writeback \\ \hline
Functional Units & 2x Int ALUs, 2x Int Vector/FP FUs \\
 & 2x Load + 1x Store \\ \hline
ReuseSensor & ScratchPad Memory of size equivalent to  \\
 & Vector RF size (768B with 48 entries) \\
 & 16B Delta Value Register\\ 
 & 4-wide instruction generator logic\\
 & Parameter Table, State History Table\\
 & Rename Map Backup Table\\\hline
Cache & 64KB 4-Way LRU L1-I/L1-D \\
 & 256KB 8-Way LRU L2 + Stride Prefetcher \\ \hline
DRAM & 8GB Dual-Channel DDR3-1600 8x8 \\ \hline
\end{tabular}
\end{table}

\textbf{Baseline}: The baseline simulations utilize ARMNN \cite{arm_nn} with the ARM Compute Library \cite{arm_CL} as the backend. We specifically use the \texttt{CpuAcc} backend mode to ensure the utilization of CPU SVE kernels for processing the layers. This configuration employs the QAsymm8 quantization scheme, which quantizes the weights and inputs using 8-bit symmetric quantization.

The Compute Library kernels employed by the ARMNN optimizer for such networks are based on the \texttt{sdot} instruction. This choice is because the \texttt{sdot} instruction is best suited for handling 8-bit input and weight values since the \texttt{mla} instruction, as described in Section \ref{sec:BG} , only supports values starting from 16-bit and provides lower performance. Furthermore, the optimizer-invoked \texttt{sdot} kernels adopt an output-stationary dataflow, as shown in Fig. \ref{fig:softReuse}. Finally, the ARMNN optimizer rearranges weights to a memory layout suitable for the kernel and the underlying architecture to avoid unnecessary cache misses during weight loads. It also employs tiling techniques tailored to the underlying cache configuration, optimizing DNN inference performance on CPUs.

\textbf{ReuseSense:}  In the simulations utilizing ReuseSense, we integrate ReuseSensor into the baseline CPU architecture. The network evaluation workflow is similar to the baseline, but instead of invoking the \texttt{sdot} kernel, we invoke a kernel that uses ReuseSensor (CRS instruction-based kernel). Also, to ensure the efficient utilization of ReuseSense, we modify the optimizer to re-arrange weights suitable for the kernel structure shown in Fig. \ref{fig:CIKernels} when invoking CRS-based kernels.

\section{Evaluation}\label{sec:ev}

This section delves into the evaluation of ReuseSense. The baseline configuration aligns with the specifications outlined in Table \ref{tab:cpuConfig} and incorporates an optimized \texttt{sdot} kernel sourced from ARMNN, which is essentially an enhanced iteration of the kernel described in Fig. \ref{fig:softReuse}-A. In contrast, the evaluation extends to configuration employing computation reuse that uses the kernel represented in Fig. \ref{fig:CIKernels}-B, referred to as 'ReuseSense'. 

\subsection{Performance and Energy Analysis}

\begin{figure}
  \centering
  \includegraphics[width=1\linewidth]{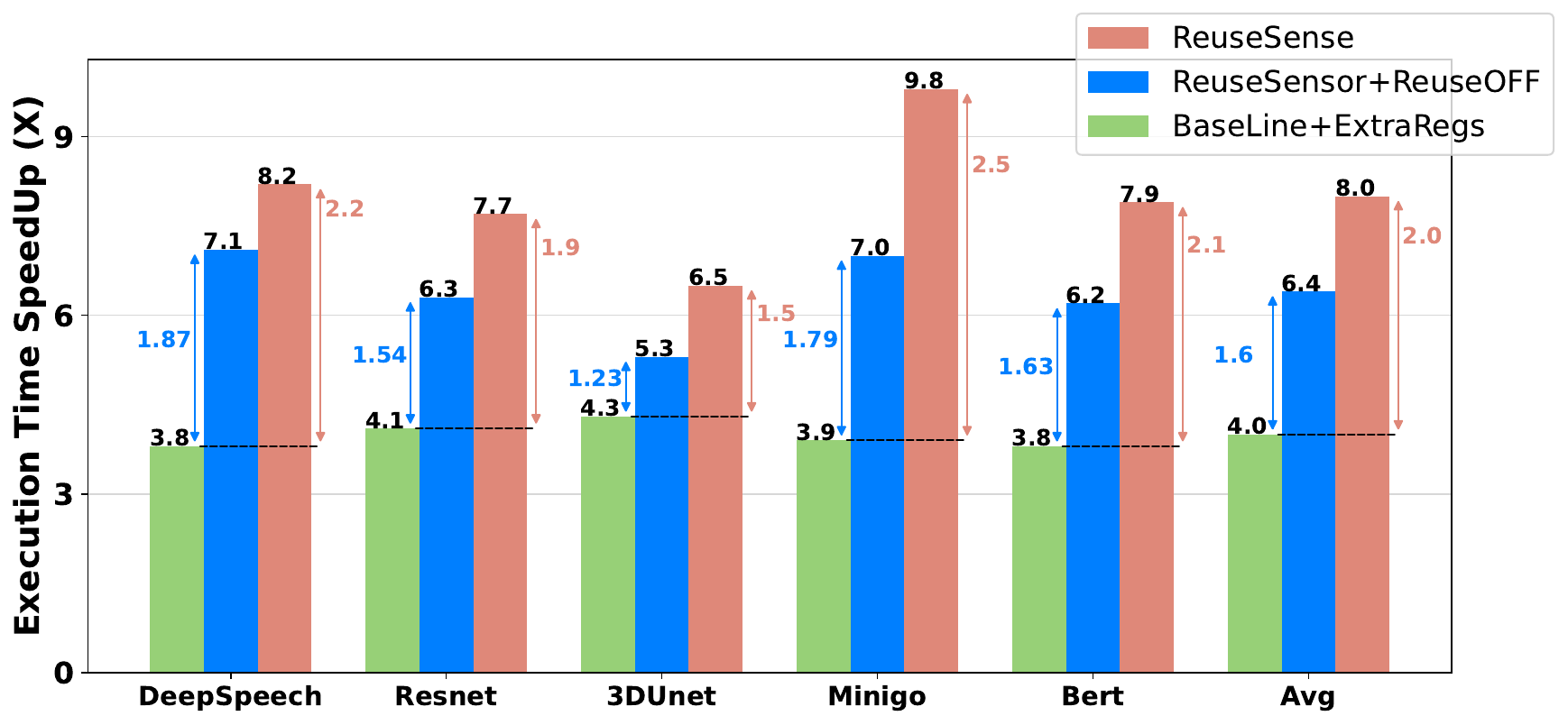}
  \caption{Speedup comparing the baseline with ReuseSense and ReuseSense+ReuseOFF. Also, we compare it with an implementation that uses a baseline with extra physical registers. }
  \label{fig:perfExecTime-sdotNoExtraRegs}
\end{figure}

Fig.~\ref{fig:perfExecTime-sdotNoExtraRegs} shows the speedup achieved by ReuseSense compared to the baseline architecture for each network in our benchmark. In this regard, 3DUnet achieves a speedup of 6.5x for ReuseSense, which is smaller than in other networks due to the large number of channels in most of its convolution layers, resulting in GEMM-ed convolutions with a considerable input size but a relatively small output size. Also, this network employs an input stationary dataflow, leading to many input loop iterations compared to the output loop iterations. As a result, there is an increase in overhead, and the benefits of reusing computations are diminished. In contrast, Minigo gains the highest speedup for ReuseSense. This is attributed to the fact that GEMMed convolution layers in Minigo have a larger output size than a relatively smaller input size, which is advantageous to the employed kernel structure. Furthermore, the additional gains for ReuseSense are a direct consequence of high computation reuse and skipping instructions due to the high input similarity in the network. Finally, ReuseSense yields an average speedup of 8x.

To demonstrate the impact of computation reuse, we contrast the performance with and without reuse implementation. In the absence of reuse (using the 'ReuseSensor+ReuseOFF' configuration and the kernel depicted in Fig.~\ref{fig:CIKernels}-A), the average speedup is 6.4x compared to the baseline. This highlights a 20\% overall improvement achieved by ReuseSense over ReuseSensor+ReuseOFF.
Furthermore, to illustrate the benefits of effectively using the extra storage requirement required by ReuseSense (i.e.,  scratchpad memory),  
we evaluate a baseline configuration with additional physical registers equal to the scratchpad size used in the ReuseSensor. As depicted in Fig.~\ref{fig:perfExecTime-sdotNoExtraRegs}, both ReuseSense and ReuseSensor+ReuseOFF achieved speedups of 2x, 1.6x respectively, compared to this configuration. Hence, approximately 4x of the total speedup can be attributed to the effective utilization of scratchpad memory.

\begin{figure}
  \centering
  \includegraphics[width=1\linewidth]{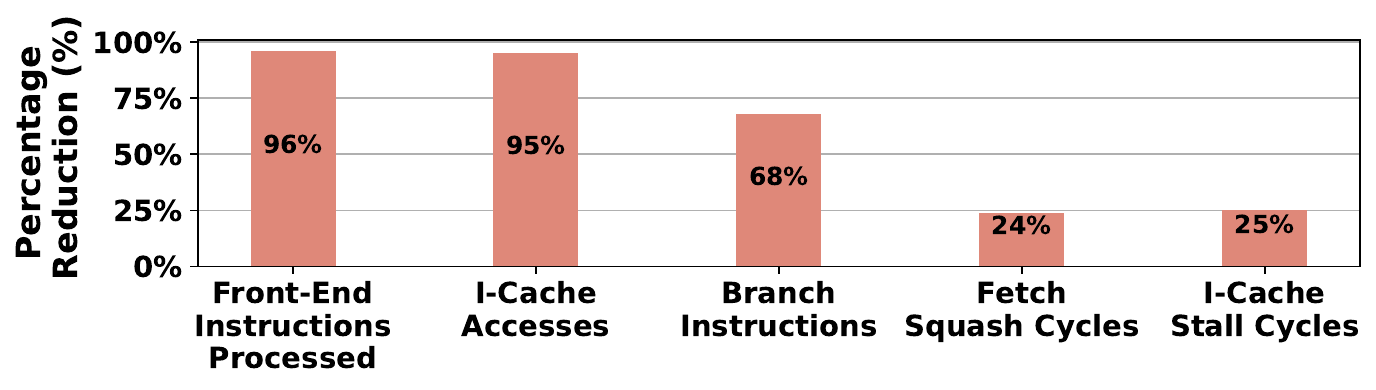}
  \caption{Percentage Reduction for different CPU hardware structures for ReuseSense compared to the baseline.}
  \label{fig:pReduct}
\end{figure}

Other factors contributing to the speedup improvements are the effective deployment of computation reuse and the efficient avoidance of significant front-end processing of instructions. During DNN inference, most processing involves invoking several kernels, and thus, by generating the memory and compute instructions required by these kernels and directly feeding them into the back end of the pipeline, ReuseSensor reduces front-end processing of instructions by 96\%, as shown in Fig.~\ref{fig:pReduct}. Also, it eliminates the need to fetch instructions from the I-Cache, leading to a 95\%  reduction in I-Cache accesses. As a result, stalls due to I-Cache are minimized. While some front-end processing is still required for handling function calls, ReuseSensor inherently avoids generating branch instructions required by loops in the kernels, leading to a 67\% reduction in branches and the associated overhead. As a result, the number of squash cycles is reduced by 23.9\%, contributing to the overall speedup achieved by ReuseSensor.

\begin{figure}
  \centering
  \includegraphics[width=1\linewidth]{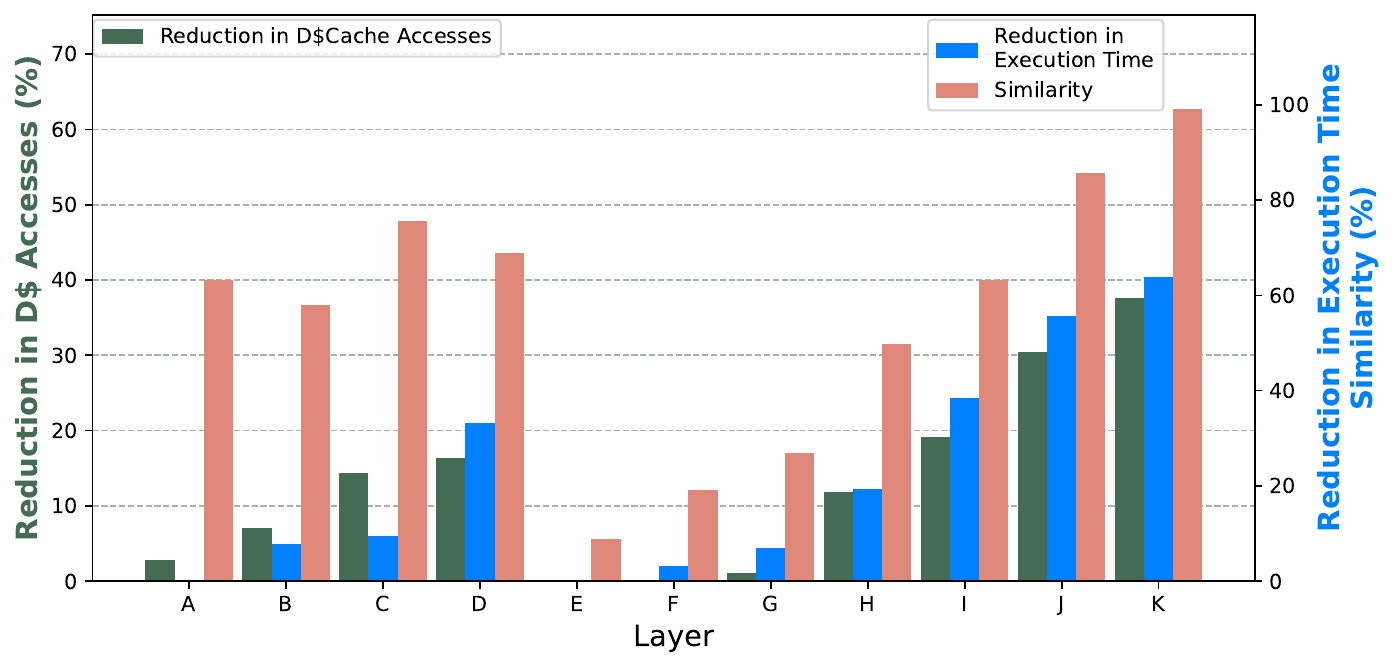}
  \caption{Comparing ReuseSense to ReuseSensor+ReuseOFF. Layers A-K are a representative pool of layers across all the DNN layers used in Table-\ref{tab:dnnSummary}.}
  \label{fig:normalVsReuseSense}
\end{figure}

We now compare the performance of ReuseSense against  ReuseSensor+ReuseOFF to understand their respective impacts.
Fig. \ref{fig:normalVsReuseSense} demonstrates the reduction in execution time achieved by utilizing the ReuseSense approach compared to ReuseSensor+ReuseOFF, for a representative set of layers across various DNN workloads used for evaluation in this work (Table \ref{tab:dnnSummary}). It also provides insights into the input similarity of each layer, along with the decrease in Data cache accesses. Layers A-D have a small output size and relatively large input size, while layers E-K have similar input and output sizes or larger output sizes. For layers with very low input similarity, ReuseSense does not yield a significant improvement and may even cause a slowdown due to the overhead of loading previous inputs and computing delta without being able to skip many compute or weight load instructions. However, as input similarity gradually increases, the percentage of instructions skipped and the reduction in data cache accesses increases, reducing execution time for ReuseSense. However, even if the input similarity is high for small layers, we see little gains in execution time due to overhead incurred by ReuseSensor kernels. In contrast, for larger layers, we see that with the increase in input similarity, there are higher reductions in data cache accesses and execution time. 
It is important to note that 100\% input similarity does not translate to a 100\% decrease in execution time, as ReuseSensor still needs to generate other instructions, such as input, previous input, and output loads, delta computation instructions, and output stores, even if all weight load and compute instructions are skipped. This is evident in the case of layer K, which shows a 60\% reduction in execution time despite having 99\% input similarity. Nevertheless, percentage input similarity translates to the same percentage reduction in the number of generated weight load and computation instructions by design.

\begin{figure}
  \centering
  \includegraphics[width=1\linewidth]{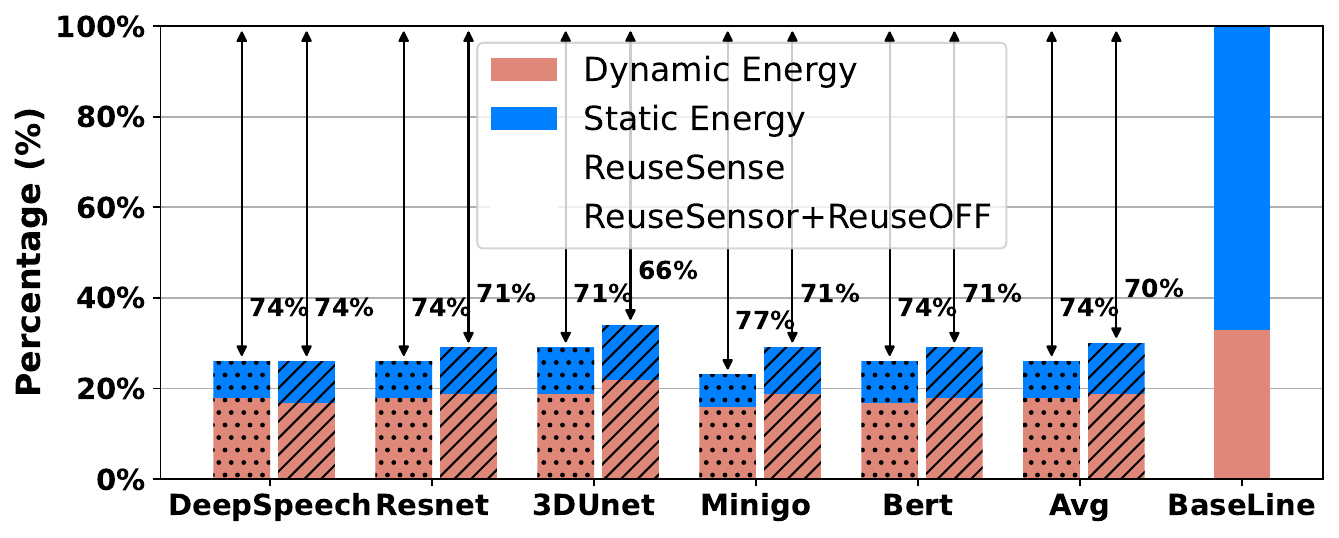}
  \caption{Reduction in total energy consumption compared to the baseline for ReuseSense and ReuseSensor+ReuseOFF.}
  \label{fig:energy}
\end{figure}

Fig. \ref{fig:energy} illustrates the total energy consumption reduction achieved by ReuseSense and ReuseSensor+ReuseOFF compared to the baseline. Across the networks, ReuseSense achieves an average reduction of 74\%, while ReuseSensor+ReuseOFF achieves a reduced average reduction of 70\%. There is an overall reduction in dynamic energy consumption of 47\% and 42\% for  ReuseSense and ReuseSensor+ReuseOFF, respectively, and the remainder of the benefits in energy come from reducing execution time, which decreases static energy.

\begin{figure}
  \centering
  \includegraphics[width=1\linewidth]{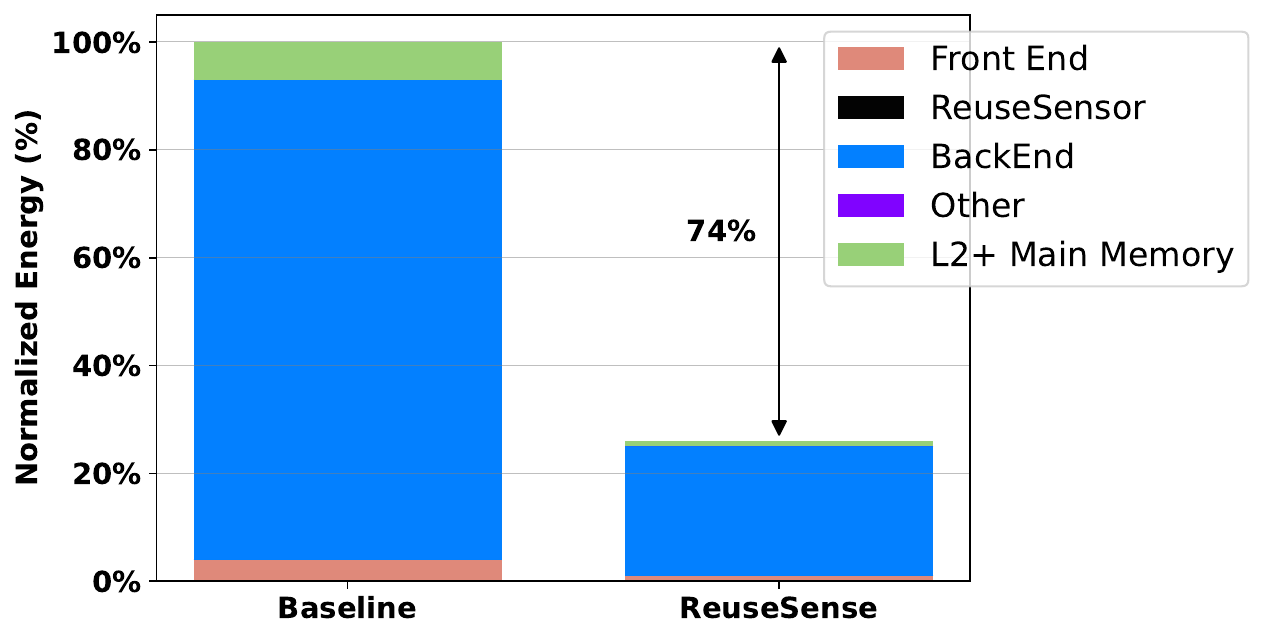}
  \caption{Total Energy (Dynamic+Static) consumption breakdown for the baseline and  ReuseSense.}
  \label{fig:energyBreakDown}
\end{figure}

To gain further insights into the energy distribution within the processor, Fig. \ref{fig:energyBreakDown} shows the average percentage of energy consumed by the main components of the architecture for all the networks. Notably, the configuration with ReuseSense enhances energy savings by avoiding redundant computations and loads, which is reflected in lower energy in the backend and the L2+Memory groups. Furthermore, it also slightly reduces the front-end energy consumption due to reducing the number of instruction fetches.

\subsection{Hardware Overhead Analysis}

ReuseSensor, as shown in Table \ref{tab:cpuConfig}, comprises various components that enable its efficient operation. The key components include a 768B on-chip scratchpad memory, a parameter table, and a delta value register. Additionally, it incorporates an
instruction generation logic that generates and sends instructions directly to the backend of the pipeline. Collectively, these components contribute to a total memory footprint of approximately 868B for ReuseSensor. Despite its additional functionality, our proposal remains lightweight, needing a small footprint of less than 1KB. The efficient utilization of underlying Out-of-Order (OoO) structures allows ReuseSense to achieve its goals while minimizing its memory requirements. Furthermore, our analysis reveals that deploying ReuseSense incurs an area overhead of less than 0.05\% compared to the baseline processor.

\section{Related Work}
\label{sec:related}

Related works to ReuseSense can be grouped into the two main categories described below. 

\subsection{DNN Performance Improvement on CPUs}

Several works have focused on enhancing DNN performance on CPUs; in this regard, SAVE \cite{SAVE} incorporates a sparsity-aware vector engine that intelligently skips ineffectual computations resulting from sparsity, reducing computational overhead. Similarly, SparCE \cite{sparCE} adopts HW/SW co-design techniques using ISA extensions to skip redundant code blocks caused by sparsity. SparseDNN \cite{SparseDNN} uses kernel-level and network-level optimizations catered for sparse networks, and Sparse CNN \cite{SparseCNN} uses an efficient sparse matrix multiplication algorithm to improve sparse DNN and CNN inference on CPUs.
ZCOMP \cite{ZComp} addresses the cross-layer memory footprint of DNNs by utilizing vector load-store compression techniques. 
On the other hand, REDUCT \cite{Reduct}, strategically employs ISA extensions and places lightweight tensor compute units near caches. This design minimizes data movement and bypasses the OoO stage processing, improving overall performance. NIOT \cite{NIOT} specifically targets the inference of Transformers on modern CPUs by deploying an optimized framework tailor-made for Transformer execution. In addition to these specific approaches, various co-optimizations and techniques have been explored to improve DNN processing performance on  CPUs. A comprehensive survey of these techniques can be found in the work by Sparsh et al. \cite{DNNForCPUSurvey}.

\subsection{Computation Reuse and  Similarity in DNNs}

Numerous techniques in the literature have effectively harnessed the concept of similarity in Deep Neural Networks (DNNs) through various approaches. Riera et al. \cite{inputSimilarityMarc} focus on computation reuse for DNN inference, implementing their method in a custom accelerator. Servais et al. \cite{adaptiveComputationReuse} take a similar path but tailor their approach for CNN training on tensor-core-based accelerators, thereby optimizing training processes. On the contrary, Deep Reuse \cite{deepReuse} groups similar neuron vectors into clusters and utilizes cluster centroids to exploit computation reuse, effectively accelerating CNN inferences. Adaptive Deep Reuse \cite{AdaptiveDeepReuse} extends the concept further by dynamically adjusting the degree of reuse to exploit input similarity during CNN training. Both works implement their strategies within the Tensorflow framework at the software level, allowing evaluation on GPUs.
In contrast, MERCURY \cite{MERCURY} employs a Random Projection with Quantization \cite{RPQ} to detect and leverage input similarity, resulting in accelerated DNN training for FPGA-based hardware accelerators. Additionally, techniques such as CREW \cite{CREW}, SumMerge \cite{SumMerge}, and UCNN \cite{UCNN} explore computation reuse in the dimension of weight repetition within DNNs.

Many of the aforementioned techniques heavily depend on specific software frameworks or specialized accelerators to harness computation reuse effectively. However, directly deploying such approaches on general-purpose CPUs would necessitate substantial modifications to the core structures of modern OoO processors or might not yield sufficient effectiveness when implemented as software solutions. In contrast, ReuseSense is framework-independent and directly exploits computation reuse at the core level, effectively utilizing existing CPU resources with minimal additional structures for orchestration support.

\section{Conclusions}
\label{sec:conc}

In this work, we introduce ReuseSense, a hardware scheme leveraging input similarity to efficiently exploit computation reuse for DNN inference on CPUs. Our contributions include evaluating input similarity across various DNN models and showcasing that input similarity exists even in cases where inputs are not part of a sequence. Also, we show that a software-only approach to exploiting computation reuse in CPUs is ineffective. In response, we propose ReuseSense, a hardware scheme that leverages input similarity to avoid executing ineffectual instructions. It employs a ReuseSensor, a hardware structure that autonomously generates the instructions needed to evaluate a DNN kernel and skips them when it senses that an input value is equal to a previous one. We implement ReuseSense on a state-of-the-art ARM CPU and show its effectiveness in decreasing energy consumption and improving performance. Compared to the baseline, ReuseSense achieves an average speedup of 8x while decreasing the total energy consumption by 74\% on average.
% \pagebreak

\section*{Acknowledgments}
\label{sec:ack}

This work has been supported by the CoCoUnit ERC Advanced Grant of the EU’s Horizon 2020 program (grant No 833057), the Spanish State Research Agency (MCIN/AEI) under grant PID2020-113172RB-I00, and the ICREA Academia program. We sincerely thank Diya Joseph for her perennial support to this work since its inception.

%%%%%%%%%%%%%%%%%%%%%%%%%%%%%%%%%%%%

\bibliographystyle{IEEEtranS}
\bibliography{refs,medelyRefs}

\end{document}